\documentclass[12pt]{iopart}
\usepackage{epsf}
\usepackage{epsfig}
\usepackage{graphicx}
\newcommand{\beq}{\begin{equation}}
\newcommand{\eeq}{\end{equation}}
\newcommand{\beqa}{\begin{eqnarray}}
\newcommand{\eeqa}{\end{eqnarray}}
\newcommand{\md}{\mbox{d}}
\def\simgt{\rlap{\lower 3.5 pt \hbox{$\mathchar \sim$}} \raise 1pt \hbox {$>$}}
\def\simlt{\rlap{\lower 3.5 pt \hbox{$\mathchar \sim$}} \raise 1pt \hbox {$<$}}
\def\sz{\scriptsize}
\newcommand{\cpc}[3]{Comput.\ Phys.\ Commun.\ {\bf #1} (19#2) #3}
\newcommand{\epjc}[3]{Euro.\ Phys.\ J.\ {\bf C#1} (19#2) #3}
\newcommand{\xjpg}[3]{J.\ Phys.\ {\bf G#1} (19#2) #3}
\newcommand{\zpc}[3]{Z.\ Phys.\ {\bf C#1} (19#2) #3}
\newcommand{\plb}[3]{Phys.\ Lett.\ {\bf B#1} (19#2) #3}

\newcommand{\npb}[3]{Nucl.\ Phys.\ {\bf B#1} (19#2) #3}
\newcommand{\prd}[3]{Phys.\ Rev.\ {\bf D#1} (19#2) #3}
\newcommand{\prl}[3]{Phys.\ Rev.\ Lett.\ {\bf #1} (19#2) #3}

\newcommand{\annu}[3]{Annu.\ Rev.\ Nucl.\ Part.\ Sci.\ {\bf #1} (19#2) #3} 
\catcode`\@=11
\def\@citex[#1]#2{\if@filesw\immediate\write\@auxout{\string\citation{#2}}\fi
  \def\@citea{}\@cite{\@for\@citeb:=#2\do
    {\@citea\def\@citea{,\penalty\@m}\@ifundefined
       {b@\@citeb}{{\bf ?}\@warning
       {Citation `\@citeb' on page \thepage \space undefined}}%
\hbox{\csname b@\@citeb\endcsname}}}{#1}}
\def\citer{\@ifnextchar [{\@tempswatrue\@citexr}{\@tempswafalse\@citexr[]}}
\def\@citexr[#1]#2{\if@filesw\immediate\write\@auxout{\string\citation{#2}}\fi
  \def\@citea{}\@cite{\@for\@citeb:=#2\do
    {\@citea\def\@citea{--\penalty\@m}\@ifundefined
       {b@\@citeb}{{\bf ?}\@warning
       {Citation `\@citeb' on page \thepage \space undefined}}%
\hbox{\csname b@\@citeb\endcsname}}}{#1}}
\catcode`@=12
\begin{document}
\begin{flushright}
 CERN-TH/99-30\\
 UCL/HEP 99-03\\
 January 1999
\end{flushright}
\title[]{Heavy-Flavour Production at HERA\footnote[7]{Summary of the 
working group on `Heavy Flavour Physics' at the 3rd UK Phenomenology
Workshop on HERA Physics, Durham, UK, 20 -- 25 September 1998. To
appear in the proceedings.}}
\author{M E Hayes\dag\ and M Kr\"amer\ddag}
\address{\dag\ University College London, Physics and Astronomy Dept., 
 London, U.K.}
\address{\ddag\ CERN, Theoretical Physics Division, CH-1211 Geneva 23, 
                Switzerland}
\begin{abstract}
We review the theoretical and experimental status of heavy-flavour
production at HERA. The results presented include some outstanding
issues in charm and beauty photoproduction, charm production in DIS
and quarkonium production.
\end{abstract}


\section{Introduction}
The production of heavy flavours (charm and beauty) in high energy
$ep$ collisions at \mbox{HERA} provides new opportunities to study the
dynamics of perturbative QCD and to extract information on the proton
and photon structure.

The heavy-flavour cross section at \mbox{HERA} is dominated by
photoproduction events where the electron is scattered by a small
angle, producing photons of almost zero virtuality.  The dynamics of
charm and beauty photoproduction can be probed at \mbox{HERA} in a
wide kinematical region, the available centre-of-mass energy being
more than one order of magnitude higher than in fixed-target
experiments. Resolved-photon interactions, where the photon fluctuates
into parton constituents, which undergo hard scattering with the
partons from the proton, contribute significantly at \mbox{HERA}
energies and allow one to study the photon structure. Because of the
high cross sections in photoproduction, some sort of tagging must be
done to allow the data to be recorded to tape in the experiment. A
charm meson may be tagged and a jet finder may also be used, with a
certain number of high transverse energy jets demanded. In the first
case information on the charm meson is available. In the second case
the observables of the meson, the jet and the correlation between the
two may be examined. This sheer wealth of variables and currently
available statistics is what makes photoproduction most interesting.
All the previous studies from normal photoproduction may be repeated
with a charm tag, but thought should be given to original analyses.

As a complement, heavy-flavour production at \mbox{HERA} can be
studied in deep inelastic scattering (DIS). The finding of a scattered
positron and the distribution of these events in bins of the parton
momentum fraction $x$ and the photon virtuality $Q^2$ is well
documented in the literature. With the additional requirement of a
tagged charm meson (typically $D^*$) a way is found to measure the
charm tagged content of the proton structure function $F_2$
($F_2^c$). Since $F_2^c$ is evolved from the gluon content, it
provides a measurement of the gluon density in the proton. Large
statistics in the normal $F_2$ measurement and the added measurement
of $F_2^c$ constrain the theory greatly. Inclusive $F_2$ is predicted
to contain between 10\% and 25\% charm and it must therefore be
properly taken into account in global analyses of structure function
data. Differential cross sections for $D^*$ production in DIS have
also become available and allow a more detailed study of the charm
production mechanisms than is possible with the inclusive $F_2^c$
measurement alone.

The production of heavy-quark bound states has been the subject of
intense study during the past few years. Exciting phenomenological
developments followed from the application of non-relativistic QCD
(NRQCD), an effective theory that disentangles physics on the scale of
the heavy-quark mass, relevant to the production of a heavy-quark
pair, from physics on the scale of the bound state's binding energy,
relevant to the formation of the quarkonium. The NRQCD approach
implies that so-called colour-octet processes, in which the
heavy-quark--antiquark pair is produced at short distances in a
colour-octet state and subsequently evolves non-perturbatively into a
physical quarkonium, must contribute to the cross section. Although
gluon fragmentation into colour-octet quark pairs appears as the most
plausible explanation of the large direct $\psi$ production cross
section observed at the Tevatron, NRQCD factorization is still not
definitely established on a quantitative level. The analyses of
inclusive charmonium production at \mbox{HERA} offer unique
possibilities to test general features of the quarkonium production
mechanisms and to establish the phenomenological significance of
colour-octet processes.

In the following we will discuss various aspects of charm and beauty
production at \mbox{HERA}, organized into the three above-mentioned
areas: photoproduction, deep inelastic scattering and quarkonium
production. Additional information may be found in
\citer{Cole98,Thorne98}.

\section{Photoproduction}
The photon structure is being probed in new kinematic regimes at
HERA. Tagging of charmed mesons or semileptonic decays in
photoproduction events is providing a wealth of information on charm
in the photon and the processes involved in its production. Thanks to
the overall higher cross section for photoproduction events, there
exists a large statistics sample from the 1996 and 1997 data sets
alone (37 pb${}^{-1}$). Since photoproduction of charm and beauty is
much less affected by non-perturbative and higher-order effects than
hadroproduction cross sections, the \mbox{HERA} measurements can
provide important tests of the heavy-flavour production dynamics.

\newpage
The photon--proton cross section may be written as a sum of a direct 
and a resolved-photon contribution:
\begin{eqnarray}\label{dir-res}
\hspace{-1.5cm}\lefteqn{\sigma_{\gamma P}(Q^2) 
= \sum_{i} \int\! \md x_P f^P_i (x_P,\mu_P) 
\sigma_{i\gamma}(x_P,\alpha_s(\mu_R),\mu_P,\mu_R,\mu_\gamma,Q^2)}
\nonumber\\
\hspace{-0.5cm}&+& \sum_{ij} \int\! \md x_P \md x_\gamma f^P_i (x_P,\mu'_P) 
f^\gamma_j (x_\gamma, \mu_\gamma)
   \sigma_{ij}(x_P,x_\gamma,\alpha_s(\mu'_R),\mu'_P,\mu'_R,\mu_\gamma,Q^2),
\end{eqnarray}
where $f^P_i (x_P,\mu_P)$ and $f^\gamma_j (x_\gamma, \mu_\gamma)$
denote the density functions for parton $i$ and $j$ in the proton and
in the photon, respectively. The renormalization scale is labelled
$\mu_R$, and $Q^2$ is the hard scale of the interaction. The
factorization scales $\mu_P$ and $\mu_\gamma$ are related to the
subtraction of divergences associated with the collinear emission of
partons from the incoming parton in the proton and in the incoming
photon, respectively.

\begin{figure}[htb]
\vspace{5mm}
  \epsfysize=4cm
  \epsfxsize=4cm
  \centerline{\epsffile{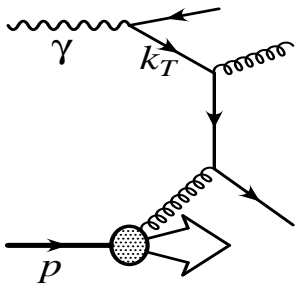}}
\caption[dummy]{\small \label{factor} 
A NLO direct photoproduction process.}
\end{figure}

The factorization of collinear initial-state singularities has to be
performed at NLO where Feynman diagrams like Figure~\ref{factor}
contribute.
Although this diagram forms part of the ${\cal O}(\alpha\alpha_s^2)$
cross section, if the virtuality $k_T^2$ of the propagator labelled in
the figure is less than $\mu_\gamma^2$, then its contribution is
already included in the LO-resolved process. Hence the separation of
direct and resolved reactions is ambiguous beyond LO and only the sum
is meaningful. This is explicitly shown in (\ref{dir-res}), where the
factorization scale $\mu_\gamma$ appears in both the direct and the
resolved contributions. Total cross sections, single-inclusive
distributions as well as heavy-quark correlations have been calculated
to next-to-leading order accuracy \cite{gpnlo}.

In higher orders, potentially large terms $\sim \alpha_s
\ln(p_T^2/m_c^2)$ arise from the collinear branching of gluons or photons
into heavy-quark pairs or from the collinear emission of gluons by a
heavy quark at large transverse momentum. For $p_T \gg m_c$ these
terms may spoil the convergence of the perturbation series and should
be resummed by absorbing the collinear
$\ln(p_T^2/m_c^2)$-singularities into heavy-quark parton densities and
fragmentation functions.  Applications of this approach to charm and
$D^*$ photoproduction at $p_T > m_c$ can be found in \cite{FF}.  For a
comparison of the different theoretical schemes, see
\cite{Cacciari-98}.

\begin{figure}[tb]
  \epsfysize=8cm
  \epsfxsize=8cm
  \centerline{\epsffile{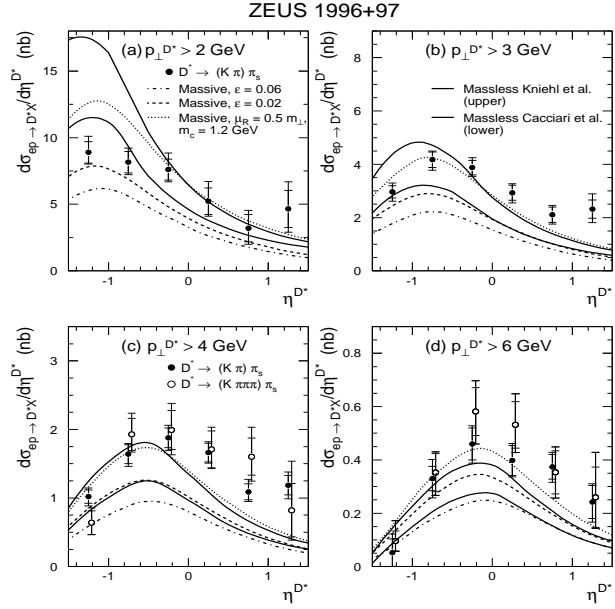}}
\caption[dummy]{\small \label{php_eta} 
The pseudorapidity distribution of the tagged $D^*$ for $p_\perp^{D^*}
> 2,3,4,6$ GeV.  The points are ZEUS data for the 1996 and 1997 data
sample \cite{zeus98}.  The inner error bars are the statistical errors
and the outer error bars are the statistical and systematic errors
added in quadrature. The curves are different theoretical NLO
predictions \cite{FMNR,FF}. The variable $\epsilon$ is the
non-perturbative parameter in the Peterson et al. fragmentation
function \cite{PSSZ83}.}
\end{figure}

\begin{figure}[tb]
  \epsfysize=8cm
  \epsfxsize=8cm
  \centerline{\epsffile{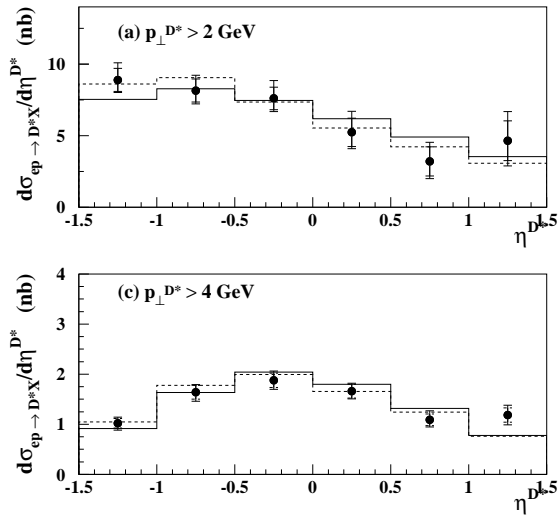}}
\caption[dummy]{\small \label{eta_xs} 
 The pseudorapidity distribution of the tagged $D^*$ for $p_\perp^{D^*}
 > 2$ GeV and $p_\perp^{D^*} > 4$ GeV.  The points are ZEUS preliminary
 data for the 1996 and 1997 data sample.  The inner error bars are
 the statistical errors and the outer error bars are the statistical
 and systematic errors added in quadrature. The solid histogram is
 HERWIG~\cite{herwig} and the dashed histogram is PYTHIA~\cite{pythia}, 
 both having been area normalised to the data.}
\end{figure}

Experimentally, the first topic to be studied, in general, is the
production of charmed mesons, which may be identified by their decay
products. Both H1 and ZEUS have shown results on the $D^*$ meson in
photoproduction \cite{zeus98,h198}. This channel is favoured because
of its clear signal. A comparison of \mbox{ZEUS} data and different
NLO theory predictions for the $D^*$ pseudorapidity distribution is
shown in Figure~\ref{php_eta}.
The data lie above the NLO theory when the $D^*$ meson travels in the
forward direction, as is also seen in the DIS case. Although the
discrepancy might not be considered very significant at present, some
interesting suggestions were made for its possible causes. They are
listed here and discussed below:
\begin{enumerate}
\item `String effects' between the proton remnant and the $c$ quark 
\cite{leif}.
\item Low-$x_\gamma$ non-perturbative gluon in the photon \cite{rohini}. 
This would give rise to more charm and is currently poorly constrained by 
LEP data.
\item Effects attributed to the Peterson et al.\ fragmentation function used.
\item More beauty than predicted by the theory.
\end{enumerate}
Item (i) can be examined by comparing the available LO Monte Carlo
programs to the data. In the fragmentation process a string would be
linked between the charm quark and the forward travelling proton
remnant. The tension in this string could be responsible for pulling
the quark (and hence the meson) forward. This effect is not
implemented in the NLO calculations, so LO Monte Carlo would be more
forward than NLO calculations. This comparison was carried out at the
recent HERA Monte Carlo Workshop \cite{leonid98} and a better
agreement is seen between the shape of the LO curves and the data (see
Figure~\ref{eta_xs}). However, there are many other effects
implemented in the Monte Carlo (including item (iii)). The process of
separating them is still ongoing.

Previously ZEUS demonstrated \cite{zeusxg} its sensitivity to the
low-$x_\gamma$ regions, by the use of the observable
\begin{equation}
x_\gamma^{\rm OBS} = \frac{\sum_{\rm jets} 
E_T^{\rm jet}e^{-\eta^{\rm jet}}}{2y E_e},
\end{equation}
where the sum is over the two jets with highest transverse energy. In
LO, $x_\gamma^{\rm OBS}$ can be interpreted as an observable related
to the fraction of the photon's momentum participating in the
hard-scale interaction. At higher $x_\gamma^{\rm OBS}$ (the usual cut
is 0.75) the direct processes dominate, and at lower $x_\gamma^{\rm
OBS}$ the resolved processes dominate. If a charm meson or
semileptonic decay is then tagged in these events, statements can be
made about the charm component of the photon. This is of interest
because charm in the photon may come from the low-$x_\gamma$
non-perturbative gluon, a parton density of the photon still badly
constrained by LEP \cite{drees95}.

Demanding a jet also provides another, even harder, scale to the
calculation. Typical experimental jet transverse energies are of the
order of 6 or 7 GeV, well above the charm mass. When $Q^2 \gg m_c^2$,
charm should behave as a light quark. At the scales probed with
present data it is unclear to what extent this is true, and the
problem arises of how to treat the intermediate $Q^2$ region. Similar
theoretical issues arise for charm production in DIS and will be
discussed in the next section.

\begin{figure}[tb]
  \epsfysize=10cm
  \epsfxsize=8cm
  \centerline{\epsffile{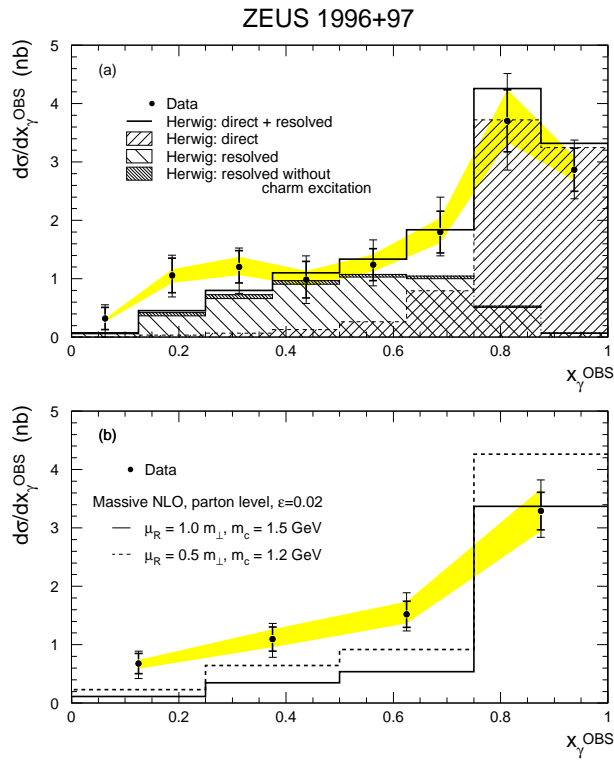}}
\caption[dummy]{\small \label{xgamma_dstar} 
The $x_\gamma^{\rm OBS}$ distribution for tagged $D^*$ decays from
charm, beauty and light quarks. The points are ZEUS preliminary data
points for the 1996 and 1997 data sample. The inner error bars are the
statistical errors and the outer error bars are the statistical and
systematic errors added in quadrature.  In the top plot the shaded
histograms show the direct and resolved contributions as predicted by
HERWIG 5.9. The solid line is the two contributions added together.
In the bottom plot the same data is shown (rebinned into four bins)
against a NLO calculation \cite{FMNR} with two choices of parameters.}
\end{figure}

ZEUS have published a measurement of $x_\gamma^{\rm OBS}$ \cite{zeus98}
with a tagged $D^*$ meson, shown in Figure~\ref{xgamma_dstar}.
This shows a clear tail to low $x_\gamma^{\rm OBS}$, which in terms of
LO Monte Carlo programs can only be explained by a large (40\%)
component of LO-resolved processes. This is heavily dependent on the
amount of charm in the parton density function used. A NLO
calculation, performed in a factorization scheme where heavy quarks
are exclusively generated at short distances from light-parton
scattering, cannot describe this long tail and falls short of both the
data and the LO Monte Carlo. This could be due to a large
non-perturbative part of the gluon in the photon, or could be
attributable to effects from the fragmentation function used.

Heavy-hadron fragmentation functions were discussed as an interesting
area of study and not just as a possible explanation as item (iii) on
the above list. The extraction of fragmentation functions from
experimental data and the value of the corresponding non-perturbative
parameters are closely related to the description of the perturbative
part of the production cross section. Different theoretical schemes
for calculating heavy-quark production require different heavy-hadron
fragmentation functions. (See \cite{ffee} for recent NLO extractions
of $D^*$ fragmentation functions from $e^+e^-$ data.)  Suggestions
have been made for defining and directly measuring fragmentation
functions at HERA \cite{mark98}.  However, the invariant mass of the
subprocess cannot be determined directly and has to be reconstructed
from observables in the final state (e.g. jet variables). It therefore
makes sense to define the fragmentation parameter in terms of the
jets. One example would be:
\begin{equation}
z = \frac{P\cdot p(D^*)}{P\cdot p({\rm jet})} = 
\frac{(E-p_z)_{D^*}}{(E-p_z)_{\rm jet}},
\end{equation}
where $P$ is the four momentum of the proton and $p(D^*)$ and $p({\rm
jet})$ are those of the meson and the jet, respectively. A modified
fragmentation function may then be defined as
\begin{equation}
D_{{\rm jet} \rightarrow D^*}(z) =\frac{1}{\sigma_0^{ {\rm jet} 
\rightarrow D^*}} \frac{d \sigma^{ {\rm jet} \rightarrow D^*}(z)}{dz}.
\end{equation}

Only a convolution of the $c \rightarrow D^*$ and $c \rightarrow {\rm
jet}$ fragmentation functions may be measured, so direct comparison
with the LEP data is impossible. With suitably large statistics, it
may be possible to do a search for scaling violations with HERA data
alone.  Work is now proceeding on the measurement of charm
fragmentation functions at HERA.

Finally, the subject of open beauty production was brought up. H1
already had released their preliminary measurement and comparison with
the LO theory \cite{beautyH1}. H1 quote, for the kinematic
range $Q^2 < 1~{\rm GeV}^2$, $0.1<y<0.8$, $p_T > 2.0$~GeV and
$35^\circ < \theta^\mu < 130^\circ$, the preliminary visible cross
section of $\sigma^{\rm vis}_{ep \rightarrow e + b\overline{b} + X} =
0.93\pm 0.08 {}^{+0.017}_{-0.07}$~nb and also quote a predicted cross
section from AROMA~2.2 \cite{aroma} of $\sigma^{\rm vis}_{ep
\rightarrow e + b\overline{b} + X} = 0.191$~nb.  This states that the
preliminary measurement of the cross section of open beauty was up to
a factor of 5 higher than the LO theory as predicted by the Monte
Carlo AROMA. It should be noted that AROMA is designed to produce only
a direct component. The magnitude of the resolved component is not
taken into account in the comparison.

At the time of the workshop, no result from ZEUS was forthcoming.
However, before going to press, a preliminary result was released of a
beauty measurement by the study of semileptonic decays to electrons
\cite{beautyzeus}. The measurement is an inclusive electron
measurement $$e^+p \rightarrow e^- + {\rm dijets} + X$$ in the
kinematic region:
\begin{itemize}
\item $Q^2 < 1~{\rm GeV}^2$, $0.2<y<0.8$.
\item Two jets with $E_T > 7,6$ GeV and $|\eta | < 2.4$.
\item An electron in the final state with $p_T > 1.6$ GeV and 
$|\eta | < 1.1$.
\end{itemize}

\begin{figure}[tb]
  \epsfysize=10cm
  \epsfxsize=10cm
  \centerline{\epsffile{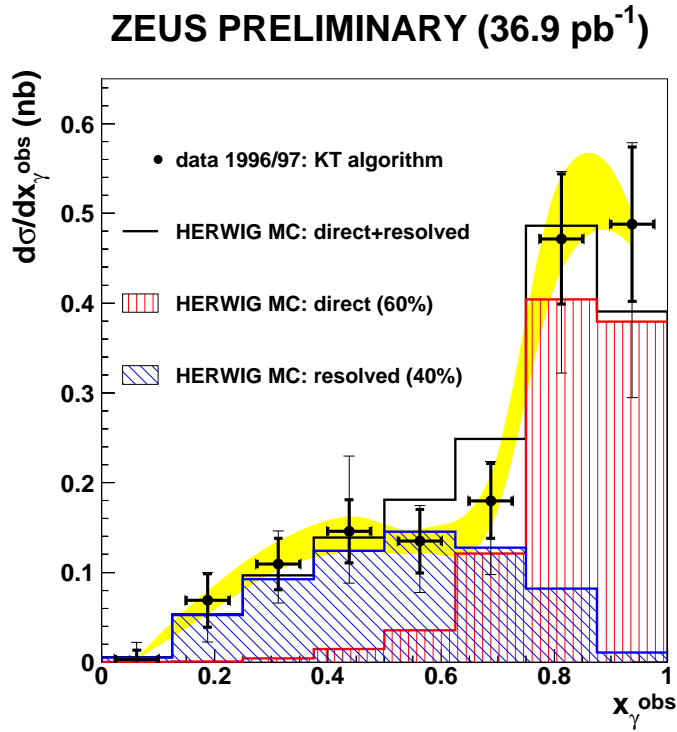}}
\caption[dummy]{\small \label{xg-sl} 
The $x_\gamma^{\rm OBS}$ distribution for tagged semileptonic decays
from charm, beauty and light quarks. The points are ZEUS preliminary
1996 and 1997 data points. The inner error bars are the statistical
errors and the outer error bars are the statistical and systematic
errors added in quadrature. The shaded histograms show the direct and
resolved contributions as predicted by HERWIG 5.9. The solid line is
the two contributions added together.  }
\end{figure}

This measurement is inclusive of all beauty, charm and light-quark
decays, but with all detector effects removed (e.g. conversion
electrons). The $x_\gamma^{\rm OBS}$ distribution (Figure~\ref{xg-sl})
shows, as with the $D^*$ study, a clear tail to low $x_\gamma^{\rm
OBS}$, which needs a significant fraction of resolved LO Monte Carlo
to describe it. The $x_\gamma^{\rm OBS}$ distribution also peaks at
high values, which is consistent with the observation of direct
processes. The preliminary fit on the fraction of the LO resolved
contribution yields $35 \pm 6$\% and agrees well with the
HERWIG~5.9~\cite{herwig} prediction of 40\%. The agreement in shape
between the data and the LO Monte Carlo is good.

\begin{figure}[tb]
  \epsfysize=10cm
  \epsfxsize=10cm
  \centerline{\epsffile{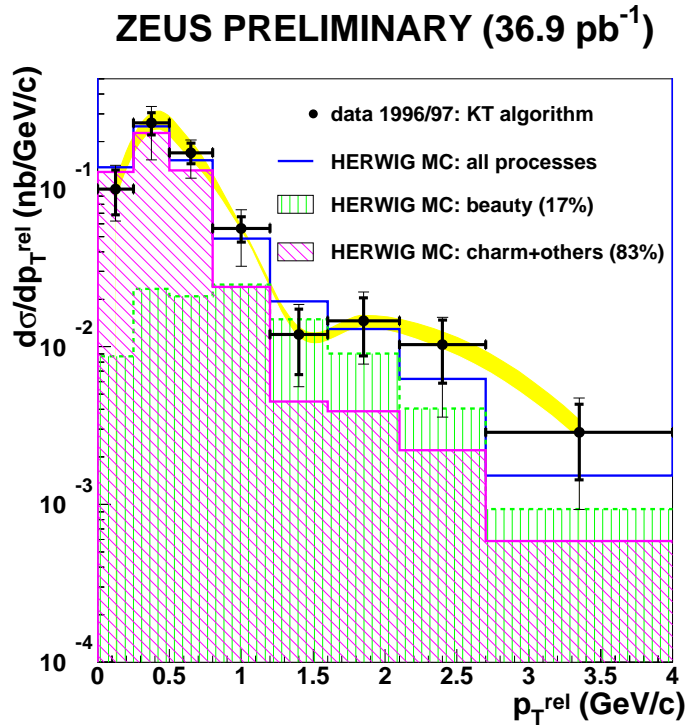}}
\caption[dummy]{\small \label{pt-sl} 
The $p_T^{\rm rel}$ distribution for tagged semileptonic decays from charm, 
beauty and light quarks. The points are ZEUS preliminary 1996 and 1997 
data points. The inner error bars are the statistical errors
 and the outer error bars are the statistical and systematic errors added 
in quadrature. The shaded histograms show the beauty and charm added to 
other components as predicted by HERWIG 5.9. The solid line is the two 
contributions added together.
}
\end{figure}

Next the $p_T^{\rm rel}$ distribution was examined. In this study
$p_T^{\rm rel}$ is the momentum of the electron transverse to the jet
axis of the closest jet (the jets were reconstructed using a
clustering algorithm, KTCLUS~\cite{ktclus}). The electrons from beauty
decays are expected to dominate at high $p_T^{\rm rel}$ because of 
the larger beauty mass. The obtained distribution is shown in
Figure~\ref{pt-sl}.  A fit was done allowing the fraction of beauty to
vary with respect to the charm and light quarks contribution (added in
the ratios given by HERWIG). A preliminary beauty fraction of $20\pm 6
^{+12}_{- 7}$\% is needed to fit the data (using the 40\% resolved
component), in agreement with the HERWIG prediction of $17$\%.

ZEUS quotes a preliminary visible cross section for beauty production
of $$\sigma^{\rm vis}_{b\overline{b}}{}(e^+p \rightarrow e^- + {\rm
dijet} + X) = 39\pm 11 ^{+23}_{-16}~\mbox{pb}$$ in the kinematic
region described above.

Finally, a preliminary overall normalization factor of 3.7 is
necessary to fit the Monte Carlo to the data.  This normalization
factor is for the complete sample and not just for the beauty
component. It should be noted that these results depend on the parton
density functions used in the analysis (for the photon
GRV-LO~\cite{GRVg} and for the proton CTEQ4~\cite{cteq4}) and also on
the charm and beauty masses used ($m_c=1.55$ GeV and $m_b=4.95$
GeV). It should also be noted that for normal dijet photoproduction a
factor of 1.8 is generally needed to normalize the Monte Carlo to the
data. In the $D^*$ with dijets study, a factor of 2.6 was necessary.

There is clear consistency between HERWIG~5.9 and the ZEUS data in the
percentages and $p_T^{\rm rel}$-distribution shapes. However, they
disagree with the large normalization factor. Taking all the caveats
in the previous paragraph into account and the different measured
kinematic regimes, it is hard to make a statement about the agreement
between the experiments at present. A comparison of the visible cross
section with a NLO calculation as well as alternative experimental
techniques using microvertex detection should allow a clarification of
the issue in the future.

\section{Charm Production in DIS}
The production of charm quarks in \mbox{DIS} has become an important
theoretical and phenomenological issue. The charm contribution to the
total structure function $F_2$ at small $x$, at HERA, is sizeable, up
to $\sim 25\%$; through the charm contribution to the scaling
violations the treatment of charm also has a significant impact on the
interpretation of the NMC data. Besides being an interesting
theoretical problem due to the presence of two hard scales, $m_c$ and
$Q$, a proper description of charm-quark production in DIS is thus
required for a global analysis of structure function data and a
precise extraction of the parton densities in the proton.

\begin{figure}[htb]
  \epsfysize=8cm
  \epsfxsize=10cm
  \centerline{\epsffile{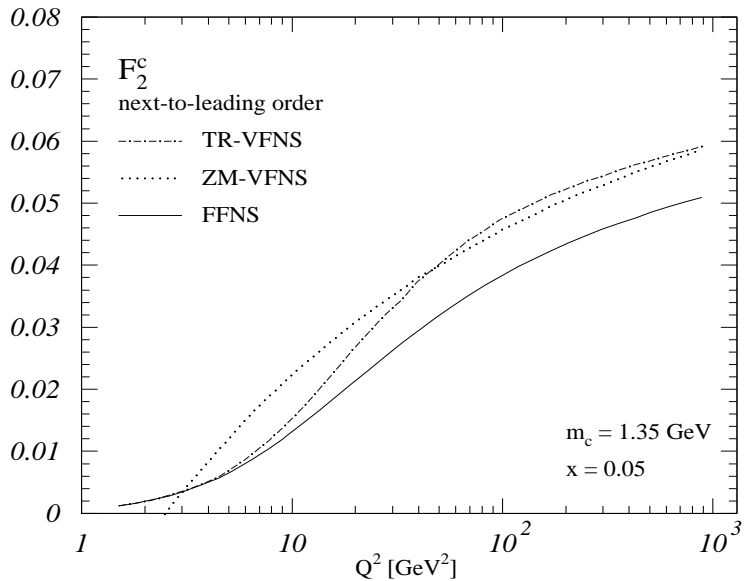}}
\vspace*{-2mm}
\caption[dummy]{\small \label{DIS-schemes} 
 Charm quark structure function, $F_c^2(x,Q^2)$ for $x=0.05$ calculated 
 in the TR-VFNS, FFNS and ZM-VFNS. See \cite{TR} for details.}
\vspace*{-2mm}
\end{figure}

At scales $Q\;\simlt\; m_c$, charm production in \mbox{DIS} is
calculated in the so-called fixed flavour number scheme (FFNS) from
hard processes initiated by light quarks ($u,d,s$) and gluons, where
all effects of the charm quark are contained in the perturbative
coefficient functions. The FFNS incorporates the correct threshold
behaviour, but for large scales, $Q\gg m_c$, the coefficient functions
in the FFNS at higher orders in $\alpha_s$ contain potentially large
logarithms $\ln^i(Q^2/m_c^2)$, which may need to be resummed. Such a
resummation can be achieved by including the heavy quark as an active
parton in the proton.  The simplest approach incorporating this idea
is the so-called zero mass variable flavour number scheme (ZM-VFNS),
where heavy quarks are treated as infinitely massive below some scale
$Q\sim m_H$ and massless above this threshold. This scheme has been
used in global fits for many years, but it has an error of ${\cal
O}(m_H^2/Q^2)$ and is not suited for quantitative analyses unless
$Q\gg m_H$. Considerable effort has been made to devise a scheme for
heavy-flavour production that interpolates between the FFNS close to
threshold and the ZM-VFNS at large $Q$. Here, we will focus on two
such schemes, the Aivaziz--Collins--Olness--Tung (ACOT) \cite{ACOT}
and the Thorne--Roberts (TR) \cite{TR} scheme, which have been used in
recent global analyses of parton distributions.\footnote[1]{For
theoretical details and the discussion of other schemes, see
\cite{Thorne98}.} These generalized VFNSs include the heavy quark as 
an active parton flavour and involve matching between the FFNS with
three active flavours and a four-flavour scheme with non-zero
heavy-quark mass. They employ the fact that the mass singularities
associated with the heavy-quark mass can be resummed into the parton
distributions without taking the limit $m_H\to 0$ in the
short-distance coefficient functions, as done in the ZM-VFNS. To all
orders (and neglecting intrinsic heavy-quark contributions
\cite{intr-HQ}) the FFNS, the ACOT scheme, and the TR scheme are
identical, but the way of ordering the perturbative expansion is not
unique and the results differ at finite order in perturbation
theory. A NLO comparison of the predictions for $F_{2}^{c}$ within the
FFNS~\cite{LRSN,HS95}, the ZM-VFNS and the TR scheme is shown in
Figure~\ref{DIS-schemes}. While the ZM-VFNS does not provide an
adequate theoretical description at low values of $Q$, the difference
between the FFNS and the TR-scheme is moderate, typically $\simlt\;
20\%$. A similar behaviour is observed for the ACOT-VFNS.

\begin{figure}[htb]
\epsfig{file=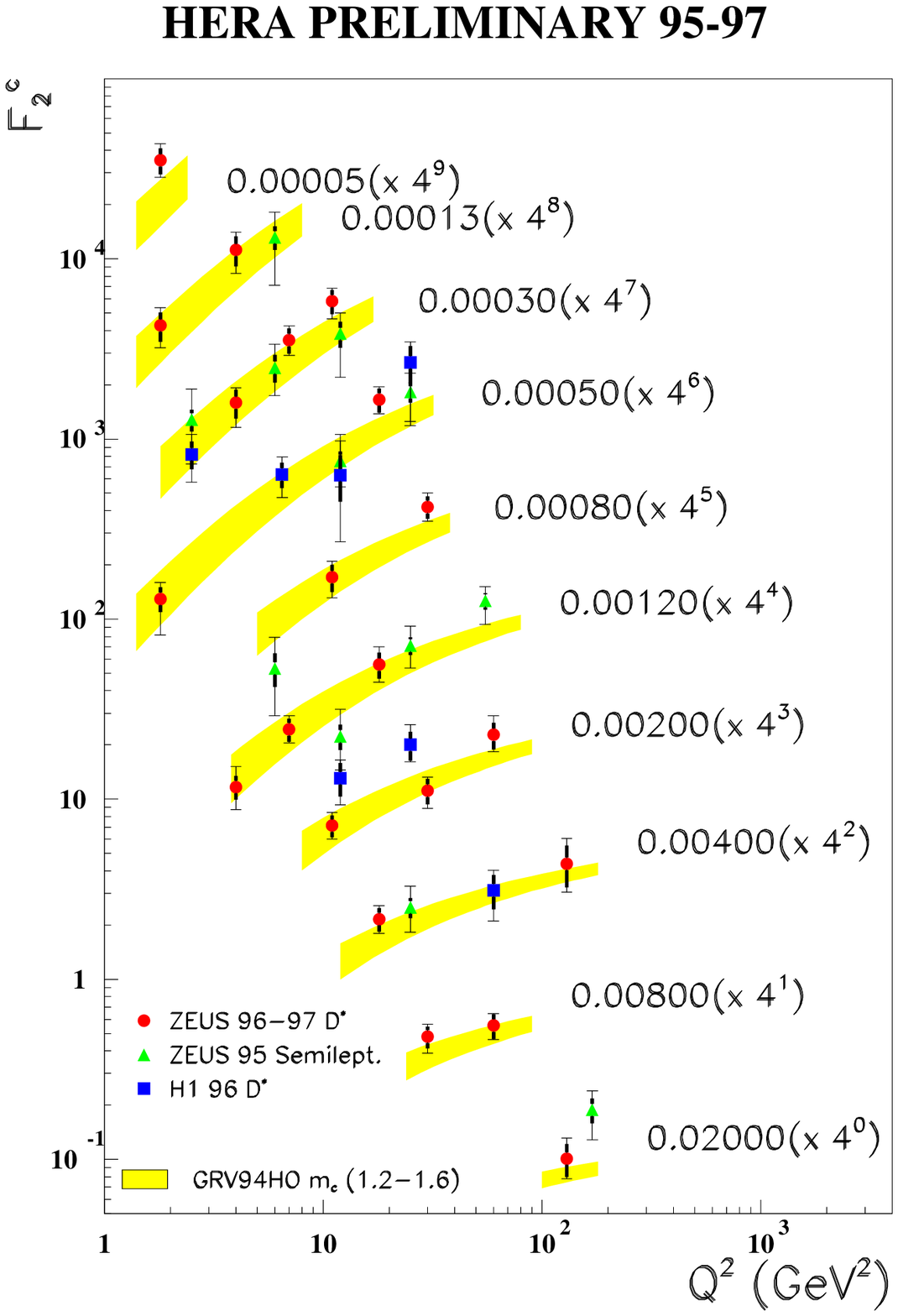,width=7.0cm,
                                bbllx=0pt,bblly=0pt,bburx=390pt,bbury=560pt}
\epsfig{file=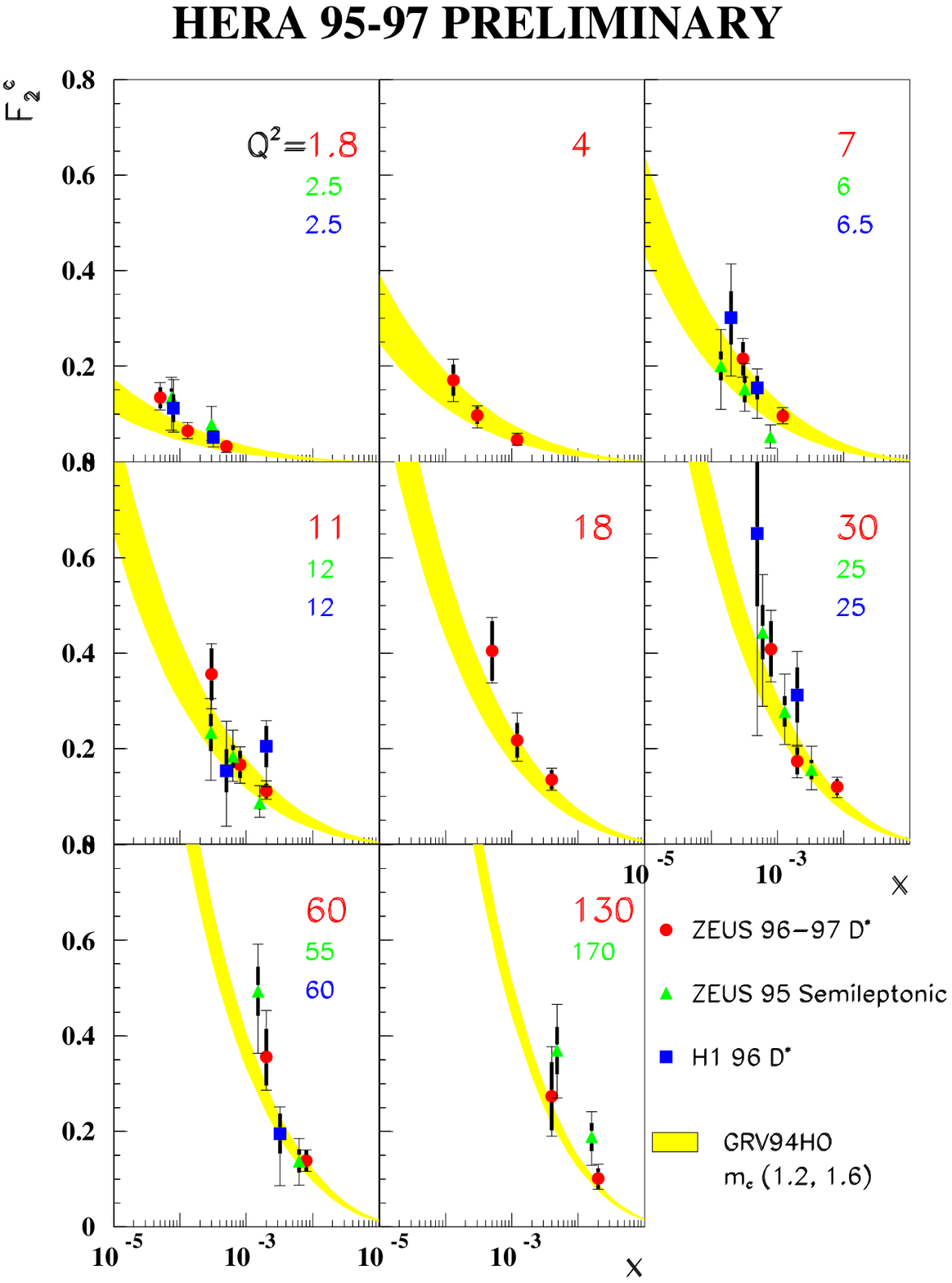,width=7.2cm,
                             bbllx=0pt,bblly=0pt,bburx=400pt,bbury=565pt}
\caption[dummy]{\small \label{F2cexp} 
 The dependence of $F_{2}^{c}$ on $Q^2$ (left) and $x$ (right)
 compared to the NLO FFNS calculation~\cite{HS95} using the GRV94
 parton densities~\cite{GRV94}.}
\vspace*{-2mm}
\end{figure}

Experimentally, ZEUS have released their $F_2^c$ measurement using
their full 1996 and 1997 data samples~\cite{ZEUS-DIS}. H1 have
released their measurement using only the 1994 to 1996 data
sample~\cite{h198}. Figure~\ref{F2cexp} shows the $Q^2$ and $x$
dependence of $F_{2}^{c}$ compared to the NLO FFNS
calculation~\cite{HS95}. The measurements are in good agreement with
the theoretical prediction. Given the experimental error and the
theoretical uncertainty due to the variation of the charm-quark mass, 
as shown in Figure~\ref{F2cexp}, present data on $F_{2}^{c}$ cannot
discriminate between the FFNS and the ACOT and TR schemes. Global
structure function fits, however, seem to favour the VFNSs over the
FFNS~\cite{Thorne98}.

The resummation of $\ln^i(Q^2/m_c^2)$ terms at higher orders in the
FFNS may not be necessary for the present analysis of neutral-current
structure functions, but it is required for a stable theoretical
prediction of the charged-current structure function $F_{3}^{c}$
\cite{BvN97}. As analysed in \cite{BvN97}, the higher-order
corrections to the lowest-order flavour excitation mechanism
$W^{\pm}+s \to c/\bar{c}$ are large and dominated by $W$-boson--gluon
fusion $W^{\pm}+g \to c/\bar{c} + \bar{s}/s$ at $x\;\simlt\; 0.1$. The
extraction of the strange quark density from the charged-current
structure functions at small $x$ thus requires a good knowledge of the
gluon density.\footnote{For a discussion of charged-current charm
electroproduction in the ACOT scheme, see \cite{KS97}.}

\begin{figure}[htb]
 \begin{center}
 \epsfig{file=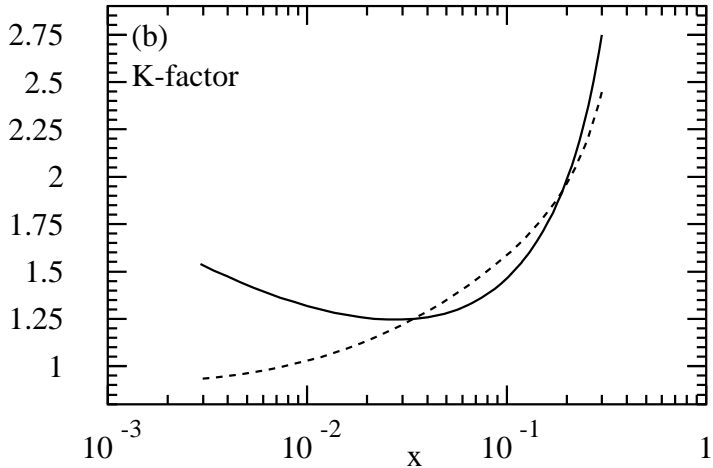,%
  bbllx=50pt,bblly=110pt,bburx=285pt,bbury=450pt,angle=270,width=10cm}
 \end{center}
 \caption[dummy]{\small \label{DIS-nnlo} 
 The $x$-dependence of the $K$-factors 
 $F_{2}^{c}(\mbox{NLO})/F_{2}^{c}(\mbox{LO})$ (solid line) and 
 $F_{2}^{c}(\mbox{NNLO})/F_{2}^{c}(\mbox{NLO})$ (dashed line). NNLO result 
 in the approximation of \cite{LM99}.}
\end{figure}

Another source of potentially large higher-order corrections to charm
production in DIS comes from the emission of soft gluons in the
threshold region. This effect has been studied within the FFNS for
charm production through photon--gluon fusion in \cite{LM99}. The fact
that heavy-quark production at HERA energies may be sensitive to
threshold effects is due to the large gluon density at small $x$,
which enhances the contribution of soft-gluon emission near the
elastic limit. An all-order resummation of threshold logarithms has
been performed in \cite{LM99} at next-to-leading logarithmic accuracy;
an approximate NNLO result has been derived for $F_{2}^{c}$ and the
single-particle inclusive distribution
$\mbox{d}F_{2}^{c}/\mbox{d}p_T$.  Figure~\ref{DIS-nnlo} shows the
effect of the approximate NNLO corrections for the inclusive structure
function. Plotted are the $K$-factors
$F_{2}^{c}(\mbox{NLO})/F_{2}^{c}(\mbox{LO})$ (solid line) and
$F_{2}^{c}(\mbox{NNLO})/F_{2}^{c}(\mbox{NLO})$ (dashed line). It can
be concluded that the soft-gluon effects are well under control for
$x\;\simlt\;0.01$, where the size of the NNLO corrections, evaluated
at the central scale $\mu = m_c$, is negligible. It has also been
shown in \cite{LM99} that the inclusion of NNLO terms stabilizes the
theoretical prediction for $F_{2}^{c}$ by reducing the scale
dependence of the NLO result.

\begin{figure}[htb]
\vspace{-3cm}
  \epsfysize=12cm
  \epsfxsize=12cm
  \centerline{\epsffile{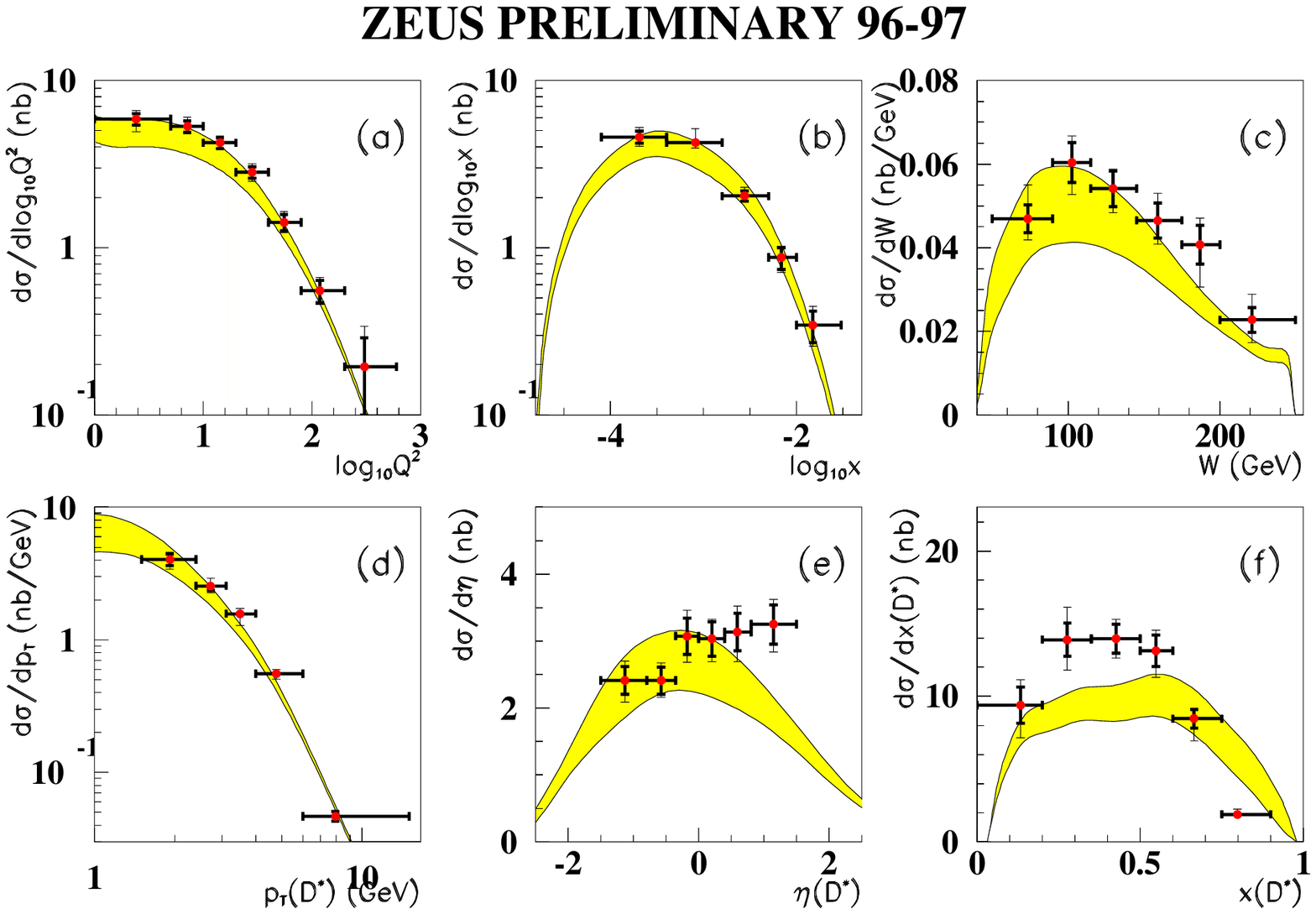}}
\vspace*{-2mm}
\caption[dummy]{\small \label{dstar_diff} 
The differential $D^*$ cross section measured by ZEUS \cite{ZEUS-DIS}
with respect to $Q^2$, $x$, the $\gamma^* p$ centre-of-mass energy
$W$, the transverse momentum and pseudorapidity of the $D^*$ and
$x(D^*)$. The shaded band shows the predictions of \cite{HS95}
including the effect of varying $m_c$ between 1.2 and 1.6~GeV.}
\end{figure}

Charm production in DIS has not only been studied for inclusive
structure functions but differential cross sections for $D^*$
production are available as well \cite{h198,ZEUS-DIS}.  Differential
distributions allow a study of the underlying production mechanism
more detailed than what is possible with the total cross section
alone.  A comparison of recent \mbox{HERA} data for differential $D^*$
cross sections in DIS with the NLO FFNS calculation~\cite{HS95} shows
good overall agreement, with a possible slight excess of data in the
forward region for large $\eta(D^*)$ and in the central $x(D^*)$
distribution (see Figure~\ref{dstar_diff}) ($x(D^*)$ is the fractional
momentum of the $D^*$ in the $\gamma^* p$ frame).  We have verified
that the excess is not driven by the small-$Q$ region of the data, so
a possible resolved photon contribution can be ruled out.
Hadronization effects may well be responsible for the deviation from
the parton-level calculation, as discussed for $D^*$ photoproduction
in the previous section.

\begin{figure}[htb]
  \epsfysize=8cm
  \epsfxsize=10cm
  \centerline{\epsffile{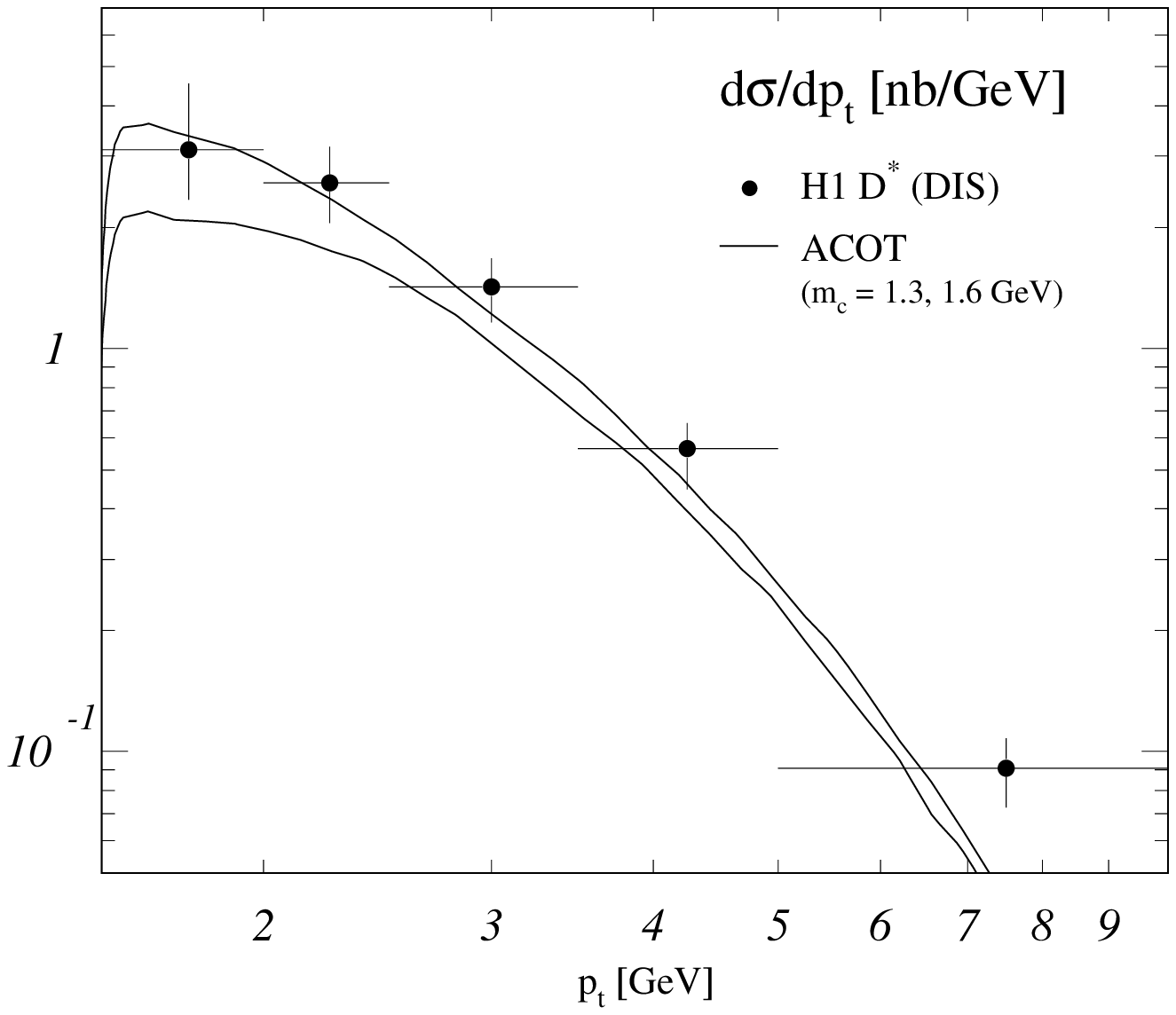}}
\vspace*{-2mm}
\caption[dummy]{\small \label{ACOT-pt} 
 Differential $D^*$ DIS cross section. Compared are recent \mbox{H1}
 data \cite{h198} with the theoretical prediction in the ACOT-VFNS
 \cite{ACOT-diff}.}
\vspace*{-2mm}
\end{figure}

Recently, new theoretical results for differential distributions in
the ACOT scheme have become available \cite{ACOT-diff}. An event
generator Monte Carlo program has been constructed, which allows to
produce arbitrary differential distributions and to impose
experimental cuts. Figure~\ref{ACOT-pt} shows a comparison between the
ACOT prediction for the $p_T$ distribution in $D^*$ DIS and recent
\mbox{H1} data \cite{h198}. Given the uncertainty in the choice of
the non-perturbative $D^*$ fragmentation function, which affects the
$p_T$ slope of the theory, the agreement is good. In
Figure~\ref{ACOT-pt} the Peterson et al.\ form \cite{PSSZ83} has been
adopted, with $\epsilon =0.078$. A smaller value for $\epsilon$, as
suggested by recent analyses of $e^+e^-$ data \cite{ffee2}, would
increase the cross section at large $p_T$ and thereby improve the
theoretical description of the data further. A more detailed
comparison between \mbox{HERA} data and VFNS predictions will be
carried out in the future.

Experimentally much effort is now being put into reducing the
systematic and statistical errors, to gain as much as possible from
the current data samples.  First, detailed systematic studies are in
progress for standard $F_2$, and these will eventually be applied to
the measurement of $F_2^c$. Some of the systematic error on $F_2^c$ is
because of the extrapolation, due to the forward excess seen by both
collaborations \cite{h198,ZEUS-DIS}.  Work is under way to implement
`string effects' described in the previous section into NLO
calculations. These appear to correct the calculation so as to agree
with the data in the forward direction.

The statistical errors may not be much improved by 1998 and 1999 data
taking, so work is under way to examine different channels. Already
ZEUS have shown a tagged semileptonic decay channel \cite{semilep},
and other channels are under investigation (e.g. $D^* \rightarrow
K\pi\pi\pi \pi_s$). Since these are statistically independent samples,
it may be possible to combine them to reduce the statistical errors.
Finally it is hoped that the new HERA luminosity upgrade planned for
next year and the inclusion of results from microvertex detectors into
the analyses will provide vastly increased statistics as well.

\section{Quarkonium Production}
The production of quarkonium at high-energy colliders has been the
subject of considerable interest during the past few years. New
experimental results from $p\bar{p}$, $ep$ and $e^+e^-$ colliders have
become available, some of which revealed dramatic shortcomings of
earlier quarkonium models (see \cite{quarkonium-reviews} for recent
reviews). In theory, progress on factorization between perturbative
and the quarkonium bound state dynamics has been made.  The
`colour-singlet model' (CSM) has been superseded by a consistent and
rigorous approach, based on non-relativistic QCD (NRQCD) \cite{BBL95},
an effective field theory that includes the so-called colour-octet
mechanisms. On the other hand, the `colour evaporation' model (CEM)
\cite{CEM-old} of the early days of quarkonium physics has been
revived \cite{CEM}.  Despite these developments, the range of
applicability of these approaches to the practical case of charmonium
is still subject to debate, as is the quantitative verification of
factorization. The problematic aspect is that, because the charmonium
mass is still not very large with respect to the QCD scale,
non-factorizable corrections may not be suppressed enough, if the
quarkonium is not part of an isolated jet, and the expansions in NRQCD
may not converge very well.  In this situation cross checks between
various processes, and predictions of observables such as quarkonium
polarization and differential cross sections, are crucial in order to
assess the importance of different quarkonium production mechanisms,
as well as the limitations of a particular theoretical approach.

According to the NRQCD formalism, inclusive quarkonium production can
be viewed as a two-step process, where the production of a heavy-quark
pair $Q\bar{Q}[n]$ in a certain angular momentum and colour state $n$
is described by a perturbatively calculable short-distance cross
section $d\hat{\sigma}_n$ and the subsequent quarkonium formation is
parametrized by a non-perturbative matrix element $\langle {\cal
O}_n^H \rangle$. The magnitude of the non-perturbative transition
probabilities is determined by the intrinsic velocity $v$ of the bound
state and the production cross section can be expressed as a double
expansion in $\alpha_s$ and $v$.

In the CEM and the related `soft colour interaction' (SCI) model
\cite{SCI}, the cross section for a specific charmonium state is given
as a fraction of the inclusive $c\overline{c}$ production cross
section integrated up to the open charm threshold. The production
fraction is assumed to be universal in the CEM, while in the SCI model
it depends somewhat on the partonic state and the possible string
configurations. In both models the colour and spin quantum numbers of
the intermediate $c\bar{c}$ state are irrelevant, and gluon emission
during hadronization is assumed to be unsuppressed.

The analysis of charmonium production in high energy $ep$ collisions
at HERA appears to be a powerful tool to test the NRQCD approach and
to discriminate NRQCD from the CEM or SCI model. The production of
charmonium at \mbox{HERA} is dominated by photon--gluon fusion, where
the photon interacts directly with a gluon from the proton.  Resolved
processes are expected to contribute significantly to the lower
end-point of the $J/\psi$ energy spectrum and might be probed at
\mbox{HERA} for the first time in the near future.  Higher-twist
phenomena such as diffractive charmonium production \cite{diffractive}
are not included in the leading-twist calculations of NRQCD and the
CEM: it can be eliminated by either a cut in the $J/\psi$ transverse
momentum $p_T\;\simgt\; 1$--$2$~GeV, or by a cut in the $J/\psi$ energy
variable $z \equiv p_p \cdot k_{\psi}\, /\, p_p \cdot
k_{\gamma}\;\simlt\; 0.9$ (in the proton rest frame, $z$ is the ratio
of the $J/\psi$ to $\gamma$ energy, $z = E_{\psi}/E_{\gamma}$).
 
Within NRQCD and at leading order in the intrinsic velocity $v$ of the
bound state, the inelastic photoproduction of $J/\psi$ comes from an
intermediate $c\bar{c}$ pair in a colour-singlet $^3\!S_1$ state and
coincides with the colour-singlet model result.  (The notation for the
angular momentum configuration is $^{2S+1}\!L_J$ with $S$, $L$ and $J$
denoting spin, orbital and total angular momentum, respectively.)
Cross sections \cite{BJ}, polar \cite{CSMpred1}, and polar and
azimuthal \cite{CSMpred2} decay angular distributions have been
calculated for the direct-photon contribution.  The angular integrated
cross section is known to next-to-leading order in $\alpha_s$
\cite{MK}. Relativistic corrections to the colour-singlet channel at
${\cal{O}}(v^2)$ were calculated in \cite{JKCW93} and were 
found to be small in the inelastic region $z\;\simlt\; 0.9$, $p_T\;
\simgt\; 1\,$~GeV.\footnote{This result has been called into question 
by a recent analysis \cite{PMM99}, which finds a sizeable ${\cal
O}(v^2)$ correction in the large-$z$ region even after a cut in the
transverse momentum, $p_T\; \simgt\; 1\,$~GeV, has been applied.}  The
colour-singlet contribution, including next-to-leading corrections in
$\alpha_s$, is known to reproduce the current $J/\psi$ photoproduction
data \cite{HERA-psi} adequately.  But there is still a considerable
amount of uncertainty in the normalization of the theoretical
prediction, which arises from the value of the charm quark mass and
the wave function at the origin, as well as the choice of parton
distribution functions and renormalization/factorization scale.

At order $v^4\sim 0.05$--$0.1$ relative to the colour-singlet
contribution, the $J/\psi$ can also be produced through the
intermediate colour-octet configurations $^3\!S_1$, $^1\!S_0$ and
$^3\!P_J$. In the inelastic region, they have been considered in
\cite{Cacciari} for the direct photon contribution and in
\cite{Cacciari2} for resolved photons, in which case the photon
participates in the hard scattering through its parton
content. Colour-octet contributions to the total photoproduction cross
section (integrated over all $z$ and $p_T$) are known to
next-to-leading order \cite{MMP97}.  The polarization of inelastically
produced $J/\psi$ due to these additional production mechanisms has
been calculated recently \cite{BKV98}.

The colour-octet production channels are kinematically different from
the colour-singlet one, because the $^1\!S_0$ and $^3\!P_J$
configurations can be produced through $t$-channel exchange of a gluon
already at lowest order in $\alpha_s$. (For the $^3\!S_1$ octet this
is true for the resolved process.) This leads to a significantly
enhanced amplitude, in particular in the large-$z$ region.  The
colour-octet contributions to $J/\psi$ photoproduction are indeed
strongly peaked at large $z$ \cite{Cacciari}. Such a shape is not
supported by the data, which at first sight could lead to a rather
stringent constraint on the octet matrix elements $\langle {\cal
O}_8[n]\rangle$, $n\in\{^1\!S_0,^3\!P_0\}$, and to an inconsistency
with the values obtained for these matrix elements from other
processes.  However, it is premature to interpret the discrepancy
between the theoretical predictions for colour-octet contributions and
the \mbox{HERA} data at large $z$ as a failure of the NRQCD approach
itself, as long as not all theoretical uncertainties that reside in
the calculation of the short-distance cross sections have been
assessed carefully. Potentially large QCD effects include higher-twist
contributions \cite{MA97}, higher-order corrections and intrinsic
transverse momentum of the partons in the proton
\cite{SMS98}.  Moreover, the peaked shape of the $z$-distribution 
is derived neglecting the energy transfer in the non-perturbative
transition $c\bar{c}[n]\to J/\psi+X$.  In reality the peak may be
considerably smeared \cite{BRW97} as a consequence of resumming
kinematically enhanced higher-order corrections in $v^2$ and no
constraint or inconsistency can be derived from the end-point
behaviour of the $z$-distribution at present. As a consequence, the
role of octet contributions to the direct process remains unclear.
The resolved photon contribution, on the other hand, could be entirely
colour-octet-dominated \cite{Cacciari2,Kniehl}. The $z$-distribution
should then begin to rise again at small $z$, if the colour-octet
matrix elements are as large as suggested by NRQCD velocity scaling
rules \cite{LMNMH92,BBL95} and fits to hadroproduction data.

The CEM and the SCI model also predict a strong colour-octet
enhancement at small and large values of $z$, again due to $t$-channel
gluon-exchange diagrams \cite{CEM2}. As in the case of the NRQCD
analysis, the impact of QCD effects (higher-twist contributions,
higher-order corrections and intrinsic transverse momentum) has to be
analysed before a final conclusion can be drawn. Since the qualitative
behaviour of the $J/\psi$ energy distribution is similar in all
existing approaches that include colour-octet contributions, it is not
well suited to discriminate between NRQCD, CEM and SCI.

\begin{figure}[tb]
 \vspace*{2.5cm}
 \begin{center}
 \hspace*{-2cm}
 \epsfig{file=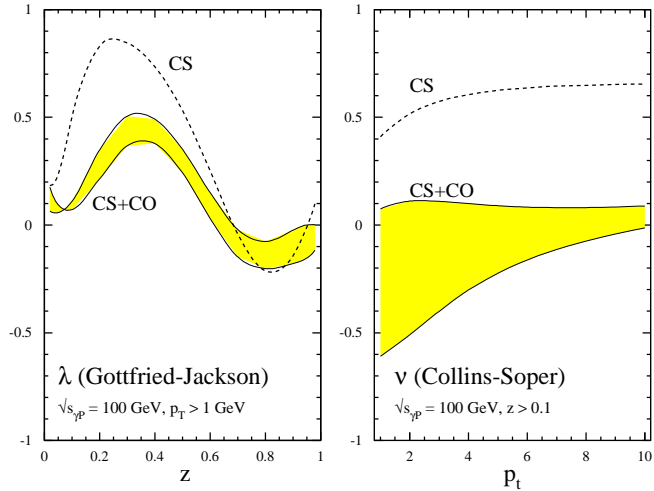,%
  bbllx=50pt,bblly=110pt,bburx=285pt,bbury=450pt,angle=0,width=4cm}
 \end{center}
 \vspace*{-2.2cm}
 \caption[dummy]{\small \label{QQ-pol} 
 Colour-singlet (CS) and colour-octet (CO) contributions to the polar
 and azimuthal angular parameters in the leptonic $J/\psi$ decay. For
 details and the definition of the Gottfried-Jackson and Collins-Soper
 reference frames see \cite{BKV98}.}
\end{figure}

The most powerful way to test NRQCD against the CEM/SCI is the
analysis of charmonium polarization. The polarization analysis in
NRQCD \cite{BR96} is based on the symmetries of the NRQCD Lagrangian,
of which spin and rotational symmetry are crucial, and the charmonium
polarization can be calculated from the spin orientation of the
intermediate $c\bar{c}$ state.  In contrast to the NRQCD approach, the
CEM/SCI predict charmonium to be produced always unpolarized. The
models assume unsuppressed gluon emission from the $c\bar{c}$ pair
during hadronization, which randomizes spin and colour. This
assumption is wrong in the heavy-quark limit where spin symmetry is at
work and soft gluon emission does not flip the heavy-quark spin.
Nonetheless, since the charm-quark mass is not very large with respect
to the QCD scale, the applicability of heavy-quark spin symmetry to
charmonium physics has to be tested by confronting the NRQCD
predictions to experimental data. Two instructive examples for
$J/\psi$ photoproduction at \mbox{HERA} are shown in
Figure~\ref{QQ-pol} \cite{BKV98}, namely the polar-angle parameter
$\lambda$ and the azimuthal-angle parameter $\nu$ in the leptonic
$J/\psi$ decay, as a function of the energy fraction $z$ and the
transverse momentum $p_T$, respectively.  The polarization signatures
predicted from colour-octet processes within NRQCD are distinctive and
can also be used to extract information on the relative weight of
colour-singlet and colour-octet contributions. Polarization signatures
have also been studied for $\Upsilon$ production in $pN$ collisions at
\mbox{HERA-B} \cite{avto}.  NRQCD predicts the $\Upsilon$ mesons to be
produced transversely polarized, again in contrast to the CEM/SCI,
which imply unpolarized quarkonium.

Another powerful way of discriminating NRQCD against the CEM is the
measurement of $\chi_c$ photoproduction. In NRQCD the
$\sigma(\chi_c)/\sigma(J/\psi)$ ratio is process-dependent and
strongly suppressed in photoproduction with respect to
hadroproduction. On the other hand, the assumption of a single
universal long-distance factor in the CEM implies a universal
$\sigma(\chi_c)/\sigma(J/\psi)$ ratio, which is not, however,
supported by the comparison of charmonium production in fixed-target
hadron collisions and early photoproduction experiments. A search for
$\chi_c$ production at \mbox{HERA}, resulting in a cross section
measurement or upper cross section limit would be crucial to settle
this issue.

\begin{figure}[tb]
   \epsfysize=14cm
   \epsfxsize=10cm
   \centerline{\epsffile{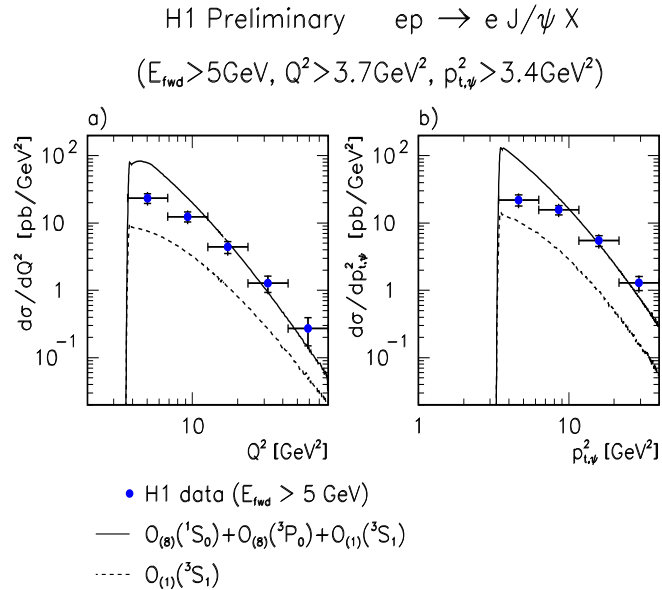}}
   \vspace*{-6cm}
\caption[dummy]{\small \label{QQ-elec} 
 Differential cross sections for $ep \to e J/\psi X$ with
 $E_{\mbox{\sz fwd}}> 5$~GeV, $Q^2 > 3.7$~GeV$^2$ and $p_T^2 >
 3.4$~GeV$^2$ \cite{H1-elec} compared to theoretical predictions
 within NRQCD \cite{FM98}.}
\vspace*{-2mm}
\end{figure}

Recently, first experimental results on inclusive $J/\psi$
electroproduction have been presented \cite{H1-elec} and compared to a
leading-order NRQCD calculation including colour-singlet as well as
colour-octet contributions \cite{FM98}. At large photon virtualities,
the analysis of $J/\psi$ leptoproduction is under better theoretical
control than photoproduction, since higher-order corrections and
higher-twist contributions should become less important as $Q^2$
increases.  Also, at sufficiently large $Q^2$, diffractive
contributions are expected to be suppressed. The experimental data for
differential cross sections in the kinematic region $Q^2>3.7$~GeV$^2$
are shown in Figure~\ref{QQ-elec}, together with the NRQCD
predictions.  The colour-singlet channel underestimates the data by a
factor $\sim 2$--$3$, but there still is a considerable amount of
uncertainty in the normalization from the value of the charm-quark
mass, the wave-function at the origin, higher-order corrections as
well as the choice of parton distribution functions and
renormalization/factorization scale. The normalization of the
theoretical prediction is significantly improved when adding
colour-octet contributions; the differential distributions, however,
seem to favour the shape predicted by the colour-singlet cross
section. More theoretical work and more experimental information
extending to larger $Q^2$ values are needed before firm conclusions
can be drawn.

Various other observables and final states can be accessed with
higher-statistics data, like photoproduction of $\psi'$ and $\Upsilon$
particles, associated $J/\psi + \gamma$ production
\cite{KR93,Cacciari2,TM97} as well as fragmentation contributions at
large $p_T$ \cite{GRS96,Kniehl}.  The future analyses at \mbox{HERA}
will offer unique possibilities to test the NRQCD approach and to
study the mechanism of quarkonium production in general.

\vspace*{-2mm}

\section*{Acknowledgements}
We wish to thank the organizers of the 3rd UK Phenomenology Workshop
on HERA Physics for creating a stimulating and pleasant atmosphere and
the heavy-flavour plenary speakers Jo Cole, Giovanni Ridolfi and
Robert Thorne for their input and enthusiasm. We would also like to
thank Matthew Wing and Mark Sutton, without whom certain results would
not have made these proceedings. We are grateful to Jim Amundson,
Matteo Cacciari, Rohini Godbole, Susanne Mohrdieck, Fred Olness, Dick
Roberts, Dave Soper and Wu-Ki Tung for valuable discussions and for
providing us with results for presentation in this summary. Finally,
many thanks go to all the participants of the workshop who have
contributed with presentations and discussions.

\section*{References}

\end{document}